\documentclass[final,3p]{elsarticle}
\biboptions{numbers,sort&compress}

\usepackage{lineno}
\usepackage[]{siunitx}
\usepackage{multirow}
\usepackage{wasysym}
\usepackage{xcolor}
\usepackage{subfigure}
\usepackage[utf8]{inputenc}
\usepackage{makecell}
\usepackage[page,toc,titletoc,title]{appendix}
\usepackage[subfigure]{tocloft}
\usepackage[section]{placeins} 
\usepackage{comment}
\usepackage{siunitx}
\usepackage{graphicx}
\usepackage{tikz}
\usetikzlibrary{circuits.ee.IEC, positioning}

\usepackage{siunitx} 
\DeclareSIUnit\sq{\ensuremath{\Box}}                           

\usepackage[rawfloats=true]{floatrow}

\usepackage{xcolor}

\usepackage{hyperref}

\journal{Nucl. Instrum. Meth. A}

\begin{document}

\begin{frontmatter}



\title{Optimization and Performance Characterization of the Second Generation Fermilab Constant Fraction Discriminator Readout ASIC}


\author[a]{Artur Apresyan\corref{1}}\ead{apresyan@fnal.gov}
\author[a]{Shuoxing Wu}
\author[a,b]{Si Xie}
\author[a]{Cristi\'an Pe\~na}
\author[a]{Tom Zimmerman}
\author[a]{Sergey Los}
\author[a]{Todd Zenger}
\author[c]{Zhenyu Ye}

\cortext[1]{Corresponding author}

\address[a]{Fermi National Accelerator Laboratory, PO Box 500, Batavia IL 60510-5011, USA}
\address[b]{California Institute of Technology, Pasadena, CA, USA}
\address[c]{Lawrence Berkeley National Laboratory, Berkeley, CA, USA}

\begin{abstract}
We present the optimization and performance characterization of the second-generation Fermilab Constant Fraction Discriminator ASIC (FCFD), designed for the readout of AC-coupled low-gain avalanche detector (LGAD) strip-sensors. The FCFD is explicitly engineered to be insensitive to signal-amplitude variations, thereby removing the need for time-walk correction that is required in other LGAD time-stamping readout ASICs. This updated version, referred to as FCFD1.1, incorporates several enhancements over the first iteration to address key challenges in AC-coupled LGAD front-end design. We outline the primary readout-ASIC design considerations for these applications, describe the methodology used to evaluate critical sensor and system parameters, and summarize the additionally implemented features. Performance measurements using injected charge signals and minimum-ionizing particles in test-beams demonstrate a time resolution of approximately 40 ps and a position resolution of roughly \SI{15}{\micro\meter} when tested with beam particles uniformly over the active area of the sensor.

\end{abstract}


\end{frontmatter}

\clearpage
\tableofcontents

\section{Introduction}

The ability to measure time-of-arrival very precisely is a critical feature of future tracking detectors and imposes a key challenge to their development. 
Many proposed future colliders such as the Muon Collider~\cite{accettura2023muon}, the FCC-hh~\cite{Sickling, Wulz277931111}, and the Electron–Ion Collider (EIC)~\cite{AbdulKhalek:2021gbh} require tracking detectors capable of simultaneously achieving precision timing and position resolutions.
Low gain avalanche diodes (LGAD) have been intensively studied as a leading candidate technology for such detectors~\cite{ACLGADprocess, 8846722, RSD_NIM, firstAC}, and in some cases have been shown to achieve 20--30~\si{\ps} performance~\cite{Apresyan:2020ipp,Heller_2022,Madrid:2022rqw,OTT2023167541,TORNAGO2021165319}.
The next key challenge is development of readout electronics capable of extracting this high precision information while being low-power and radiation hard. 

Several solutions for extracting precision timing information for LGAD sensors have been implemented for detectors at the HL-LHC, namely the ETROC and ALTIROC readout ASICs for CMS and ATLAS experiment upgrades, respectively~\cite{ETROCTed, ALTIROC2}. 
In these ASICs the a leading-edge discriminator is used for the Time-of-Arrival (ToA) measurement, in combination with a measurement of Time-over-Threshold (ToT) needed to correct for a time-walk effect. 
The ToT or signal amplitude is critically needed in order to correct for the time-walk effect, in which the ToA measurement changes sharply or ``walks'' depending on the amplitude of a signal pulse. 
The time-walk dependence can change with sensor bias voltage, temperature, or total irradiation dose due to their impact on the signal amplitude. 
Therefore, a time-walk correction needs to be customized for each channel to achieve the optimal timing resolution and continuously updated throughout the lifetime of the detector.

In a previous paper~\cite{Xie:2023flv} we presented the initial design and performance evaluation for a new approach to an application--specific integrated circuit (ASIC) for precision timing measurements: the Fermilab constant fraction discriminator (FCFD). The FCFD performs LGAD readout using the constant fraction discrimination (CFD) methodology that mitigates the time walk effect. 
The first version of this ASIC (FCFD0) was demonstrated to have 100\% signal detection efficiency, and effectively removed the need for time-walk correction. The measured contribution to the time resolution from the FCFD0 ASIC was found to be 10~ps for signals with charge above 20~fC.

In this paper we present the design of the updated version of the readout ASIC referred to as FCFD1.1, implemented in TSMC 65~nm CMOS technology. 
It is designed to read out the AC-LGAD strip sensors used in the barrel Time-of-Flight (bTOF) detector of the ePIC experiment~\cite{ABDULKHALEK2022122447} at the EIC~\cite{osti_1765663}. 
The FCFD1.1 has several major changes implemented in the design required in order provide precision time-stamping for strip AC-LGAD sensors, and to provide measurement of the signal amplitude per channel which is required for position reconstruction. 
This version is implemented as a 6-channel prototype, and provides measurement of time-of-arrival and signal amplitude for each channel. 
The ePIC bTOF detector will use 50~$\mu$m thick strip AC-LGAD sensors with 1~cm strip length and a pitch of 500~$\mu$m. 
The target performance is about 45~ps time resolution and 30~$\mu$m spatial resolution in $R$ and $\phi$ directions~\cite{Madrid:2022rqw}.

We will present the main design updates since the previous version, the methodology to precisely measure the RC-parameters of the AC-LGAD strip sensors, the specification and performance characterization of the chip using laboratory and test-beam measurements. 

\section{AC-coupled LGAD sensors}\label{sec:ACLGAD}

LGAD sensors are a class of silicon sensors that feature an extra layer of p$^+$ material, such as Boron, close to the n-p junction of traditional $\mathrm{n}^{+}$-in-p silicon sensors.
The extra layer results in a very high electric field within a depth of a few micrometers of the junction, referred to as the gain or multiplication layer. 
Electrons produced by the ionization of a minimum-ionizing particle (MIP) traversing the sensor material initiate an avalanche process in the gain layer and amplifies the signal by a factor ranging from 5 to 100. The amplification improves the signal-to-noise ratio and maintains a fast slew rate resulting in excellent timing measurement performance~\cite{Apresyan:2020ipp,Heller_2022,Madrid:2022rqw,TORNAGO2021165319,OTT2023167541}. 

In order to ensure uniformity of the multiplication along the full surface of the sensor, the pad dimensions must be far larger than the substrate thickness, resulting in an uniform electric field across the bulk material. Recent examples from the MIP Timing Detector (MTD) of the CMS experiment or the High Granularity Timing Detector (HGTD) of the ATLAS experiment have pixel dimensions of 1.3~\si{mm}$\times$1.3~\si{mm} with a substrate thickness of $35$--$50$~\si{\mu m}. 
These type of detectors uses DC-coupled LGAD (DC-LGAD), where the metal cathode is in direct contact with the silicon, collecting drifting electrons after being amplificed. 
Due to the need of a junction termination extension (JTE) to prevent pre-mature breakdowns near the metal-silicon coupling region, the pixel's metal electrode coverage is limited and a 100\% fill-factor is not achievable. 

\begin{figure}[H]
    \centering
    \includegraphics[width=0.5\textwidth]{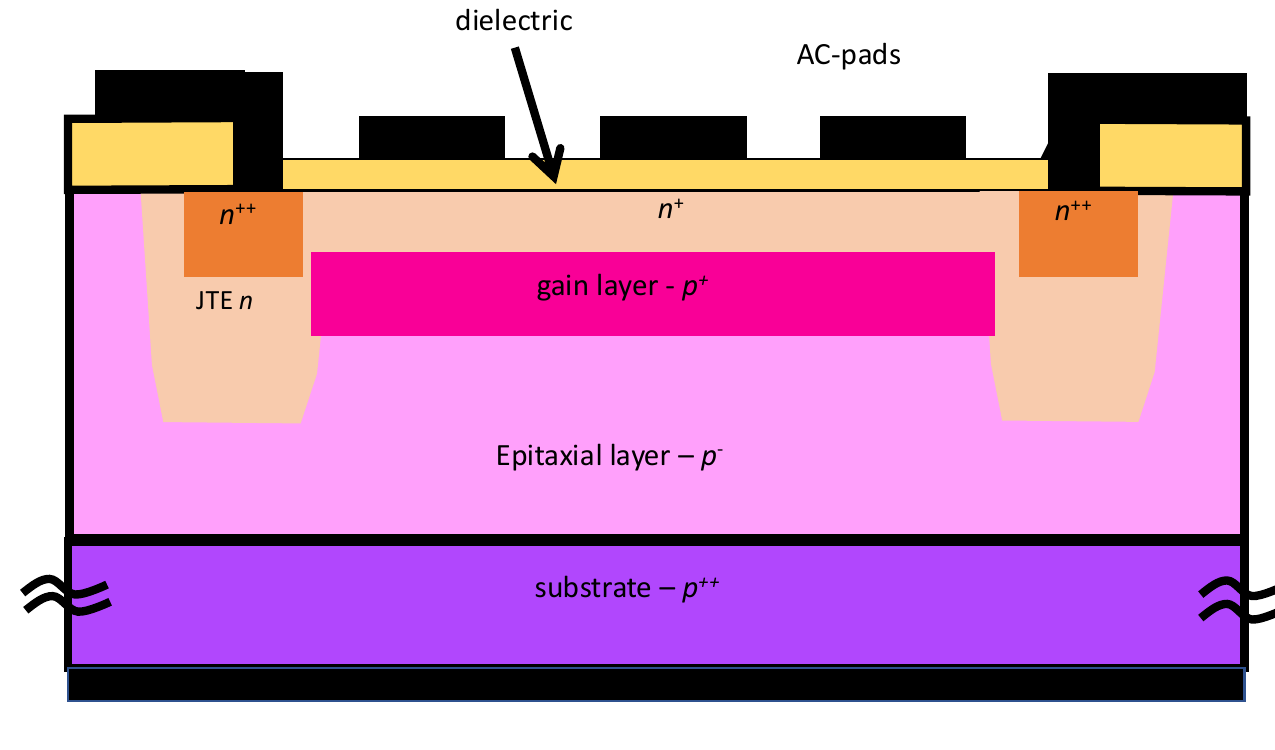}  
    \caption{A cross-sectional view of a typical AC-LGAD device.}
    \label{fig:sensor_xsec}
\end{figure}

The AC-coupled LGAD (AC-LGAD) is a novel concept to achieve a highly segmented detector while maintaining the fast timing properties. 
A cross-sectional view of a typical AC-LGAD device is shown in Fig.~\ref{fig:sensor_xsec}. 
Rather than directly coupling the metal cathodes to the silicon, the metal electrodes (AC-pad or AC-strip) are separated by a dielectric layer made of silicon oxide or nitride from the silicon. 
Drifting electrons after being amplified inside the gain layer (p$^+$) are drained to the collecting electrode placed at the edge of the whole sensor through a n$^+$ layer. 
As the electrons are drifting, signals are capacitively induced to the metal electrodes placed over the active area. 
Since we only need to place a JTE at the edge of the device to reduce the local high electrode field, the fill factor can be increased to near 100\%. 
More details of the AC-LGAD concept can be found in references~\cite{Giacomini:2019kqz,HPKsensorRef,KITA2023168009}.

\begin{figure}[H]
    \centering
    \includegraphics[width=0.5\textwidth]{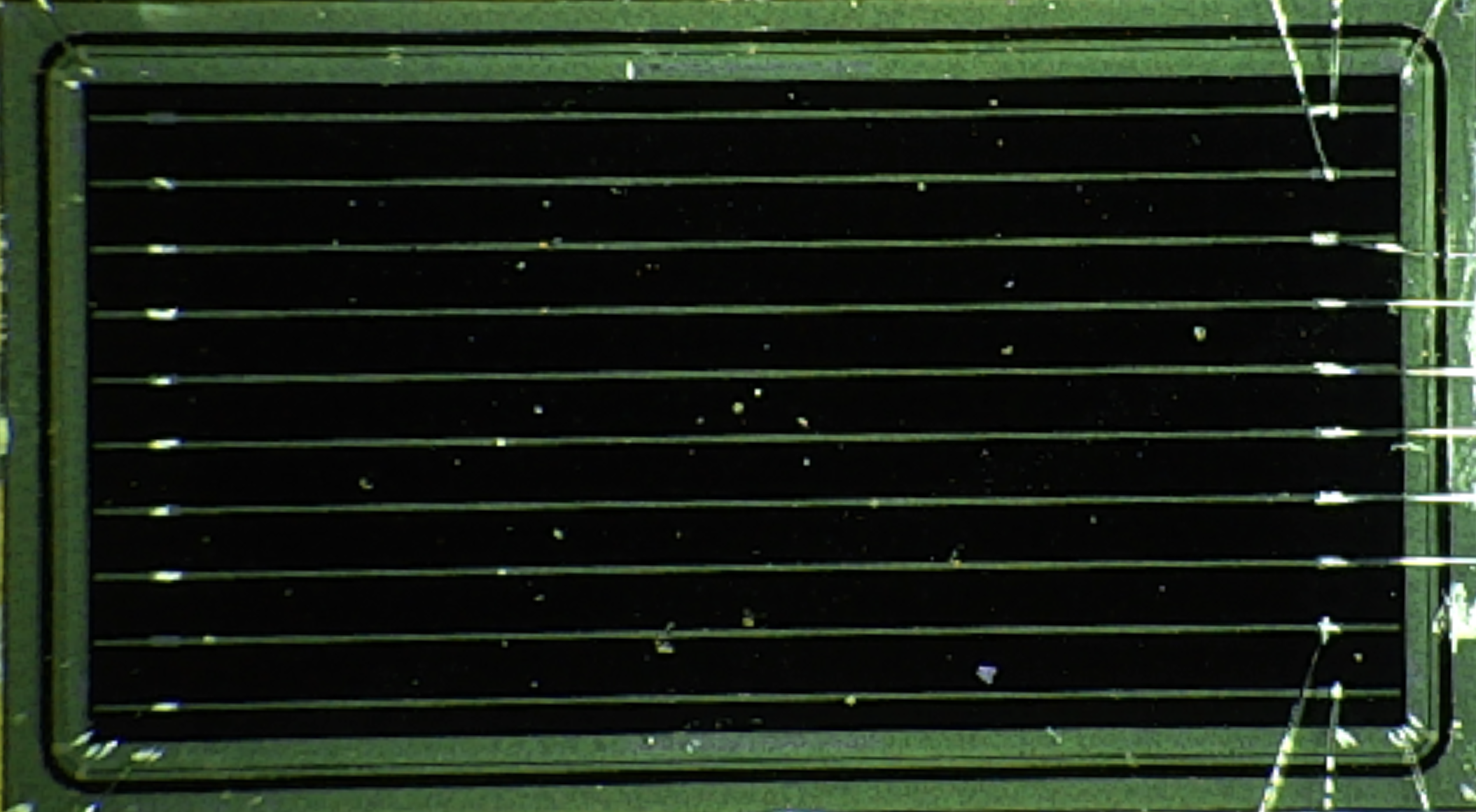}  
    \caption{Photograph of an example Hamamatsu AC-LGAD strip sensor with 1~cm strip length, 500~$\mu$m strip pitch, and 50~$\mu$m metalized electrodes.}
    \label{fig:sensorPhoto}
\end{figure}

A key feature of AC-LGAD sensor functionality is charge sharing. 
The AC-coupling of the signals results in charge being shared among neighboring electrodes. 
By using the precise amount of charge shared between the nearest and next-to-nearest neighbors, one can interpolate between the electrodes to achieve spatial resolution that is much more precise than the pitch width of the electrode strips. 
Using this method, it has been demonstrated to achieve spatial resolution performance to better than a few microns~\cite{Apresyan:2020ipp,Heller_2022,Madrid:2022rqw}. 
However, a key limitation preventing more broad use of AC-LGADs has been the lack of readout ASICs that can achieve the precise time and spatial resolution simultaneously. 

\section{Methodology for measurements of the AC-LGAD strip sensor parameters}

Precise knowledge of key parameters of the sensor is required for the design of front-end electronics, but it is often inaccessible to the ASIC designers. 
We developed a procedure to measure the RC-parameters of the AC-LGAD sensor, which we describe in this section. 
A set of electrical measurements were used to determine the capacitances, inter-strip resistances, and effective RC network behavior of the AC-LGAD sensor. 

\begin{figure}[H]
    \centering
    \includegraphics[width=0.7\textwidth]{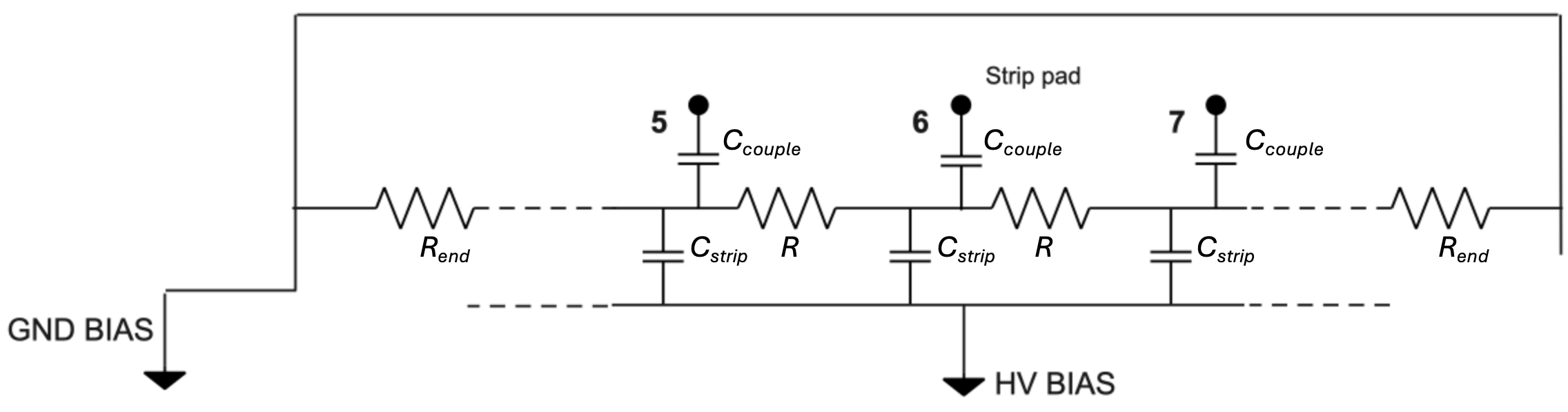}  
    \caption{The simplified RC-network model for the AC-LGAD strip-sensor. The strip pads for electrodes 5, 6 and 7 are shown. The GND\,BIAS is the bias applied to the $n^+$ resistive layer of Fig.~\ref{fig:sensor_xsec}. The $C_{\rm strip}$ is the strip capacitance to the backside (HV BIAS), and $C_{\rm couple}$ is the coupling capacitance from the strip to GND\,BIAS. The $C_{\rm couple}$ is significantly larger than $C_{\rm strip}$ to ensure that most of the charge is collected on the pad.}
    \label{fig:initialRC}
\end{figure}

In our methodology, controlled voltage-step stimuli are applied to the $n^+$ resistive layer (GND\,BIAS) electrode or individual strip pads as shown in Fig.~\ref{fig:initialRC}, and the resulting transient responses are compared with circuit simulations. 
The HV bias is always held at a fixed potential of -80~V, which ensures that the sensor is fully depleted. 
Our approach enables extraction of total strip-to-backside capacitances, distributed inter-strip resistances, and additional RC elements required to reproduce the complex time‐dependent behavior of the AC-LGAD strip network. 
All capacitance and resistances are assumed to be the same in the initial model, and are adjusted in successive steps to determine their precise values.  
The measurements are performed using a Hamamatsu sensor which contains ten one-cm-long strips, shown in Fig.~\ref{fig:sensorPhoto} and is produced with the same parameters as the SH4 sensor reported in Table 3 of ref.~\cite{Dutta:2024ugh}. 
Our measurements are performed in a few successive steps, which are described in details below. 

\subsubsection*{Step 1: Determination of the Strip Capacitance}
\noindent 
The GND\,BIAS node is bonded while all ten strip pads remain unbonded. 
A known voltage step $\Delta V_{\rm in}$ is injected through a $100~\text{pF}$ series capacitor. 
The resulting attenuated step at the GND\,BIAS node is measured.
Using the capacitive-divider relation, the total capacitance seen at the GND\,BIAS node, $C_{\rm TOTAL}$, is measured. 
Dividing by ten yields an initial estimate of the per-strip capacitance to the backside $C_{\rm strip}$:
\[
C_{\rm strip} \approx \frac{C_{\rm TOTAL}}{10} \approx 12.6~\text{pF}.
\]

\subsubsection*{Step 2: Determination of capacitance between the strip and the strip pad}
\begin{figure}[H]
    \centering
    \includegraphics[width=0.6\textwidth]{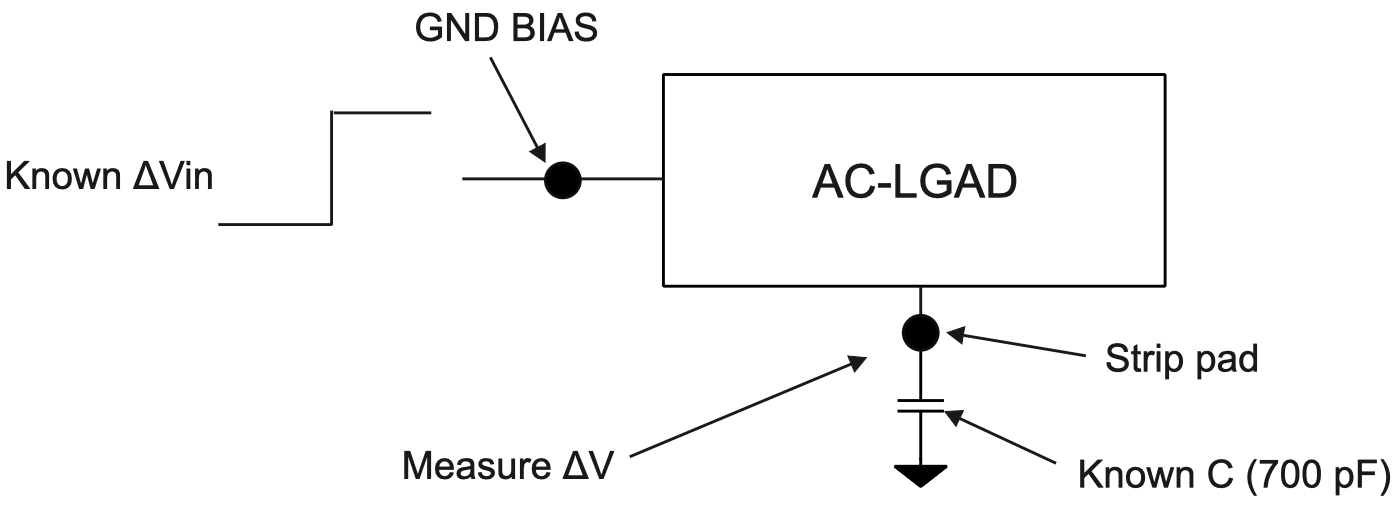}  
    \caption{A schematic diagram of the capacitive divider measuring the capacitance $C_{\rm in}$ of a single strip.}
    \label{fig:step2}
\end{figure}

\noindent
In the next step one strip pad is bonded and connected to a $700$~pF capacitor, forming a voltage divider with a coupling capacitance of $C_{\rm couple}$. 
A voltage step is applied to GND\,BIAS as shown in Fig.~\ref{fig:step2}, and the $\Delta V$ at the strip pad is measured to extract the value of $C_{\rm couple}$. The measured $C_{\rm couple}$ is the total capacitance between the strip pad and GND\,BIAS. 

\subsubsection*{Strip 3: Extraction of Inter-Strip Resistance Values}
\noindent 
In this step we measure the values of resistances shown in the diagram in Fig.~\ref{fig:initialRC}. 
The GND\,BIAS node is grounded. 
A controlled slope input $(\Delta V / \Delta t)$ is applied to one strip pad in the middle of the sensor, which injects a constant current to the RC-network at one strip. 
The induced transient voltages $\Delta V$ are then measured at all other strip inputs with a high impedance probe. 
The value of $R$ is then varied in simulations to attempt to match the observation, assuming a value of $C_{\rm couple}$ that was measured in the previous step. 
We observe that using equal inter-strip resistances does not reproduce the measured behavior, and therefore conclude that an additional resistance from each strip to the GND\,BIAS exists. 
Through simulations we determine that $R_{\rm strip\rightarrow GND} \approx 6~{\rm k}\Omega,$ is required to achieve agreement with measurements. 
Inter-strip resistances $R$ and $R_{\rm end}$ are then iteratively adjusted in simulation until convergence with experimental data is achieved.

\subsubsection*{Step 4: Network-Level RC Modeling}
\noindent
A modified electrical model of the AC-LGAD strip network now includes the coupling capacitance $C_{\rm couple}$ at each strip pad, the distributed inter-strip resistive ladder, each strip’s capacitance to the backside at HV bias, and the $R_{\rm strip\rightarrow GND}$ from each strip to GND\,BIAS. 
To verify that this modified model is complete, we now apply a step voltage to GND BIAS, leaving all strip pads open except the middle two strips (strips $5$ and $6$). 
Next, we connect strips 5 and 6 through a $50 \Omega$ coaxial cable to an oscilloscope, and compare the observed response to simulation. 
We observe that the RC-network response does not match with simulation with the rising edge being significantly steeper than what is predicted by the simulation and implying the existence of other missing components. 
Therefore, additional parasitic elements must be added to the model. 

Adding a single RC leg from each strip to ground improves agreement of simulation, but remains insufficient. 
When connecting a two-branch RC network with two distinct time constants from each strip to GND\,BIAS we find excellent agreement of simulation with observed responses.

\subsubsection*{Step 5: Additional Stimuli for Model Validation}
\noindent
To further constrain the model parameters, several complementary measurements were performed:
\begin{itemize}
    \item A $\Delta V / \Delta t$ step is applied to GND\,BIAS while strip pads $1-4$ and $7-10$ are grounded and we observe pads 5 and 6 with a FET probe.
    \item A $\Delta V / \Delta t$ step is applied to strip pad $10$ with all other pads left open and we observe all other pads with a FET probe.
    \item A $\Delta V / \Delta t$ step is applied to strip pad $6$ while all other pads are terminated in $50~\Omega$ and we observe other pads with FET probe.
\end{itemize}

These measurements yielded an independent verification of the accuracy of the model with distributed RC-elements.

\subsection{Summary of measured AC-LGAD parameters}
The combined set of measurements presented above reveals the intricate structure of the AC-LGAD strip sensors summarized below:
\begin{enumerate}
    \item The per-strip capacitance is approximately
          $C_{\rm strip} \sim 12.6~\text{pF}$.
    \item The full strip-to-GND\,BIAS capacitance measured with all strips bonded is significantly larger, $610~\text{pF}$.
    \item Accurate modeling of transient behavior requires inclusion of:
          \begin{itemize}
              \item distributed inter-strip resistances,
              \item an additional $5.8~k\Omega$ resistance to ground for
                    each strip, and
              \item a two-branch RC path from each strip to GND\,BIAS.
          \end{itemize}
    \item With these additions, simulations reproduce the observed transient
          responses for all applied stimuli, including complex multi-pad
          excitations.
\end{enumerate}

\begin{figure}[H]
    \centering
    \includegraphics[width=0.95\textwidth]{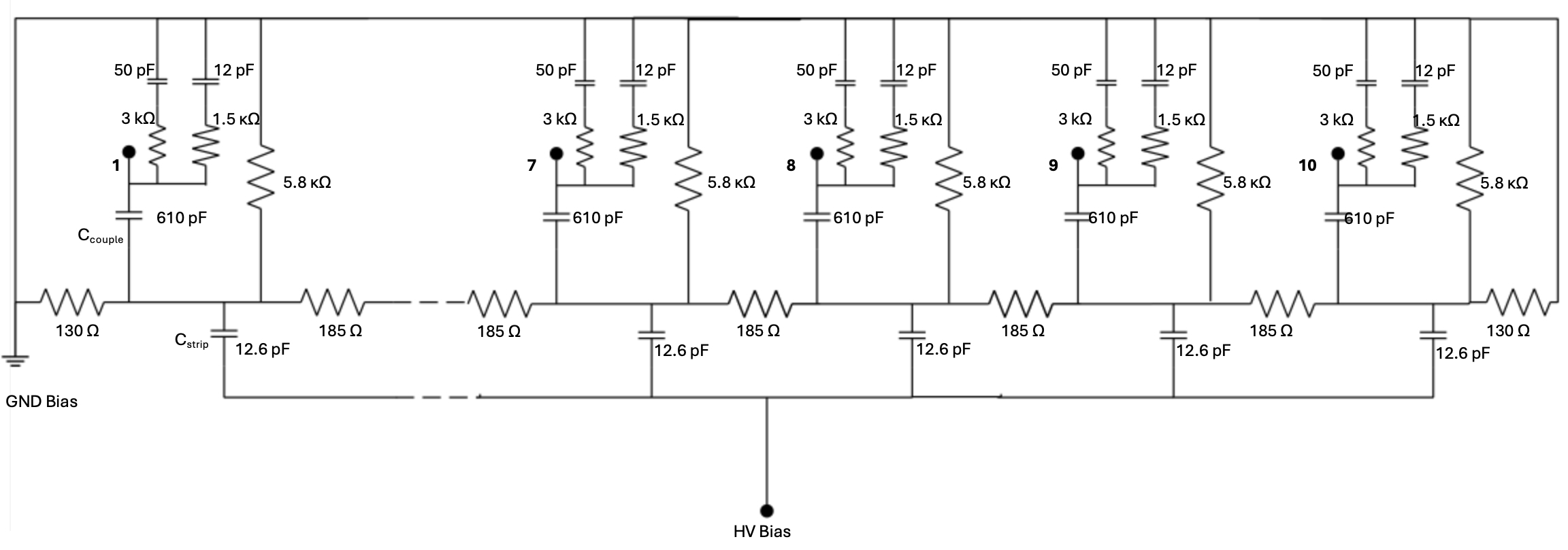}  
    \caption{The schematic diagram of the extracted model for the 1-cm long AC-LGAD strip sensor.}
    \label{fig:sensor-RC}
\end{figure}

Because the network is more complex than the simple model in Fig.~\ref{fig:initialRC} coupled, the extraction of individual parameters requires iterative adjustments. 
Measurements are repeated as needed to converge on stable values that reproduce all stimulus cases. 
The RC-network model of the sensor is shown in Fig.~\ref{fig:sensor-RC}. 
These parameters are used in the design optimization of the FCFD1.1 described in Sec.~\ref{sec:design}.

\section{FCFD1.1 Readout Chip Optimization and Amplitude Readout}\label{sec:design}

The FCFD readout ASIC implements a constant fraction discriminator (CFD) for time-stamping using the delay and zero crossing concept. 
The CFD method provide amplitude-independent timing measurement by triggering when a pulse reaches a fixed fraction of its maximum value. 
The input signal from AC-LGAD is split into two paths: a prompt path and a delayed path. 
In the prompt path, the signal is attenuated to the desired fraction, for example 50\%. 
In the delayed path, a controlled analog delay shifts the pulse in time using an RC-delay. 
The attenuated prompt pulse is then subtracted from the delayed pulse, producing a bipolar waveform whose zero-crossing occurs at the constant fraction of the original pulse amplitude. 
The discriminator detects this zero-crossing, generating a timing mark that is insensitive to variations in pulse height and ideally has no time-walk. 
A description of the FCFD design concept, details of its various components, and the characterization of its first implementation may be found in ref.~\cite{Xie:2023flv}. 

\begin{figure}[htp]
\centering
\includegraphics[width=0.5\textwidth]{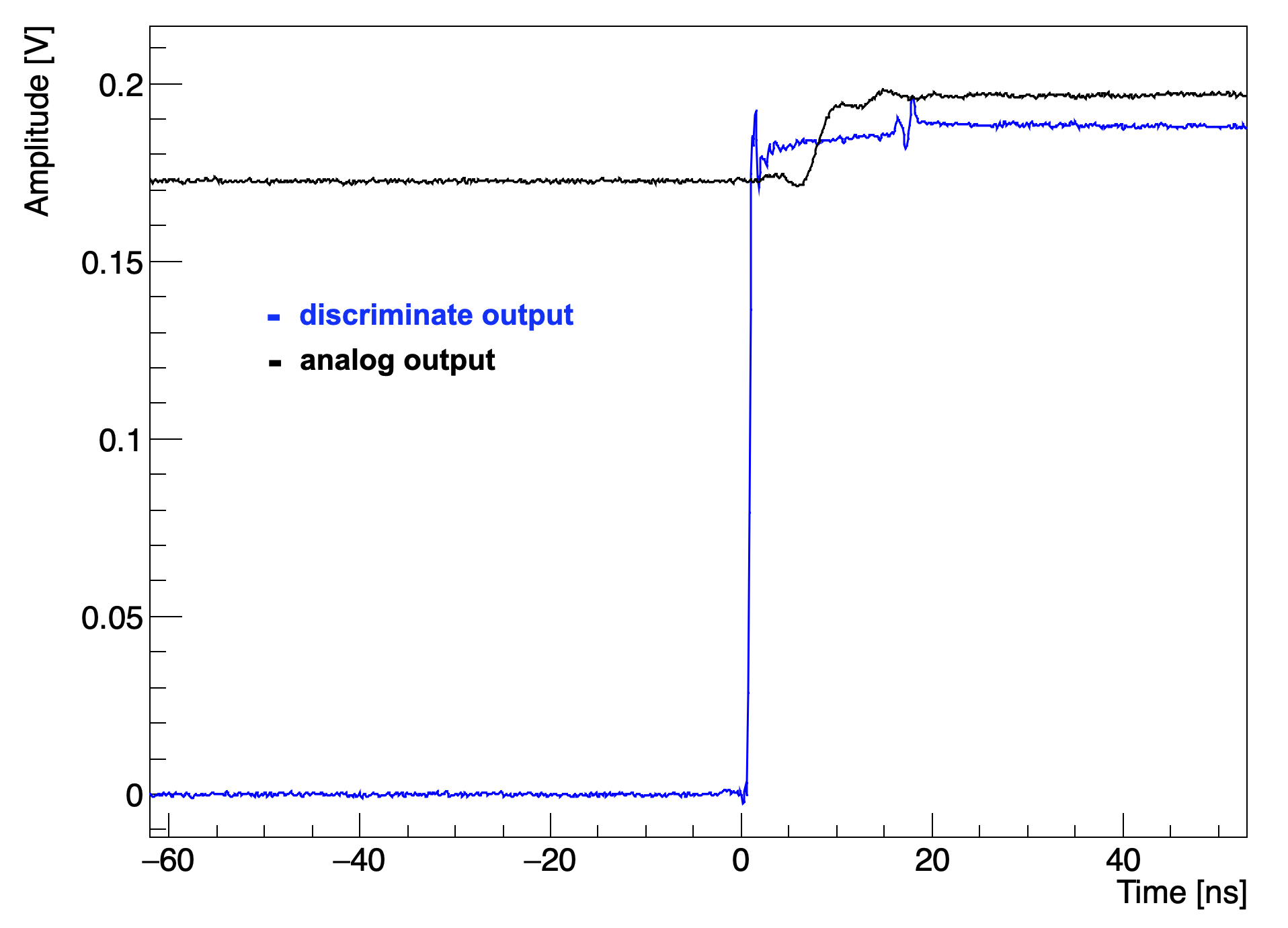}
\caption{Sample signals from the analog and discriminator output channels with the 1-cm long AC-LGAD strip sensors from testbeam data.}
\label{fig:ana_dis_output}
\end{figure}

As was demonstrated in Sec.~\ref{sec:ACLGAD}, the AC-LGAD strip-sensors constitute a complex RC-network unlike the DC-LGAD sensors, and therefore the FCFD0~\cite{Xie:2023flv} design needs to be adapted for this new application. 
The FCFD1.1 ASIC implements a number of modifications and optimizations and is specifically designed to readout AC-LGAD strip sensors with RC-parameters measured as described in Sec.~\ref{sec:ACLGAD}. 
The FCFD1.1 ASIC is designed to have dynamic range of 10-70~fC, and a low jitter of around 20~ps at the most-probable-value (MPV) of the MIP signal amplitude of $\sim$25~fC. 
The FCFD1.1 is designed to read six AC-LGAD strip channels and provides a discriminator output and a corresponding analog output for each channel, resulting in a total of six pairs of output channels. 
Example signals from the discriminator and analog outputs are shown in Fig.~\ref{fig:ana_dis_output}. 

\begin{figure}[htp]
\centering
\includegraphics[width=0.8\textwidth]{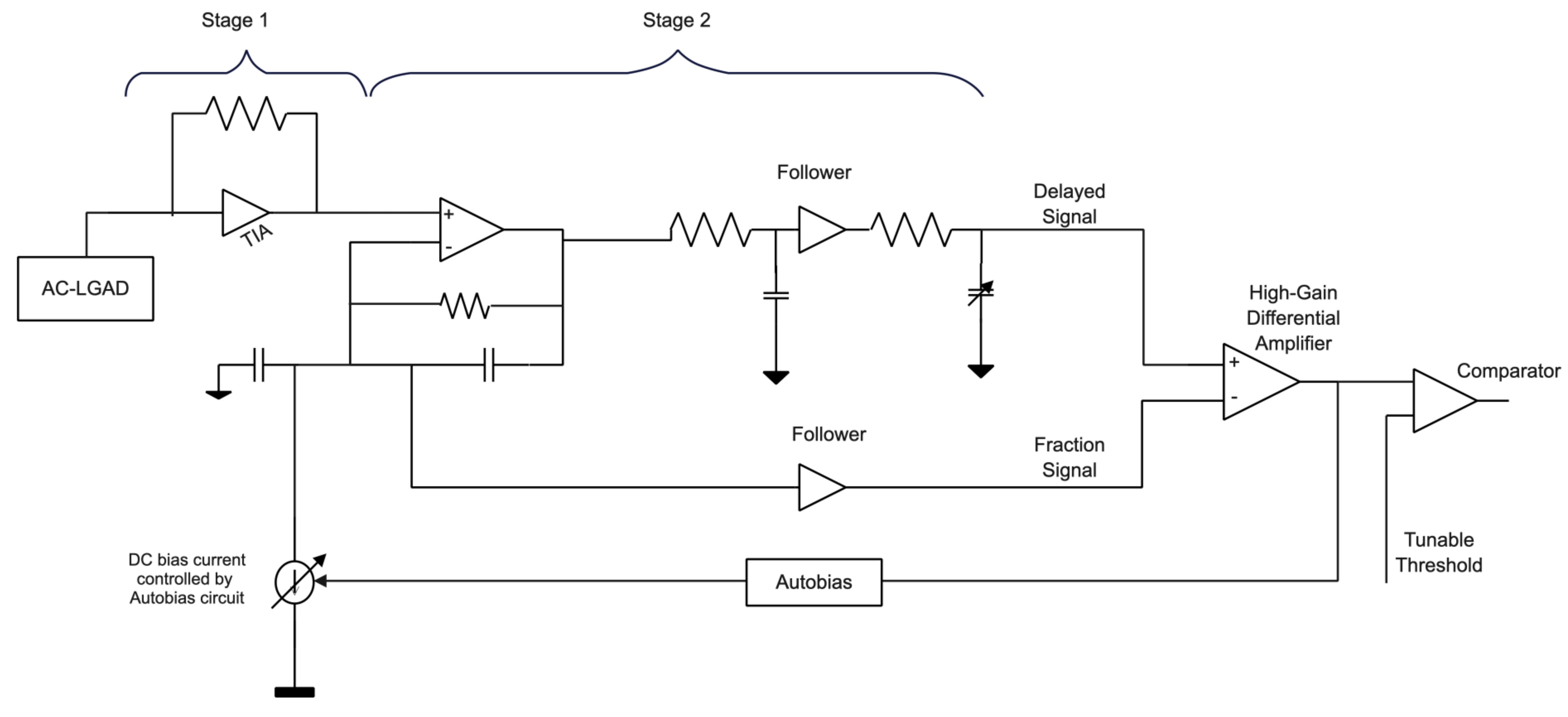}
\caption{A simplified diagram of the 2-stage amplifier used in the FCFD1.1 ASIC, and its main blocks.
\label{fig:FCFD1.1}}
\end{figure}

In the front-end of our previous CFD design, the Fraction and Delay signals that are required by the discriminator were generated using a capacitive divider in the feedback path of a single high-gain integrator stage. 
This resulted in excessive noise amplification of the AC-LGAD resistor thermal noise due to the high impedance of the capacitive feedback. 
This front end was redesigned by replacing the input integrator stage with a two-stage amplifier architecture as shown in Fig.~\ref{fig:FCFD1.1}. 
The first stage is a transimpedance amplifier (TIA) that amplifies the input current pulse and converts it to a voltage pulse by using resistive feedback. 
This effectively mitigates the noise issues observed in AC-LGAD applications. 
The second stage uses a classical non-inverting op-amp configuration with capacitive feedback to form the Fraction and Delay signals. 
The fraction is set to 50\% by using equal value feedback capacitors. 
Both signals are buffered, and the delayed path incorporates two RC-delay networks. 
The auto-bias circuit now interfaces with the second stage by supplying a variable current that flows through a relatively large resistance in the op-amp feedback loop.  
This establishes the voltage offset between the Fraction and Delay signals that is required to place the subsequent high-gain differential amplifier output at its desired operating point.

A key new feature is implemented in FCFD1.1 to enable position reconstruction using interpolation of the shared charge among neighboring electrode channels. 
An analog amplitude readout is implemented such that it appears as a DC voltage level switched to the analog output directly after the discriminator fires, and represents the amplitude of the analog signal in each channel.
This feature enables the precise spatial position reconstruction with a simple level digitization step. 
This measurement of the amplitude is extracted from the delayed signal. 
To ensure that the delayed signal is sampled at its maximum amplitude, we target the peak of the waveform, where the signal-to-noise ratio is highest. 
Because the delayed signal remains relatively fast, its peak is relatively narrow and would require precise timing to capture accurately. 
To relax this requirement, we stretch the peak by buffering the delayed signal and introducing an additional RC delay. 
A sampling switch driven by a delayed version of the discriminator output then acquires the broadened peak and stores the value on a dedicated sampling capacitor.
The stored voltage is subsequently buffered and driven onto the analog output, where it is held on the output bus for an adjustable duration of approximately 100 ns.

The arming comparator has also been redesigned and features a tunable threshold, offering more flexibility compared to the previous version. 

\section{Performance Characterization Methods and Experimental Setup}\label{sec:setup}


Similar to our previous FCFD chip studies~\cite{Xie:2023flv}, we employ several complementary methods for performance characterization.
First, we employ the internal charge injection feature of the FCFD1.1 chip to verify the timing performance in response to standardized input signals of known amplitude.
Then, we measure the response of the FCFD1.1 chip to signals produced by particles from electron and hadron beams impinging on various LGAD sensors to characterize its response to minimum-ionizing particles (MIPs). 

\begin{figure}[htp]
  \begin{center}
  \includegraphics[width=0.6\textwidth]{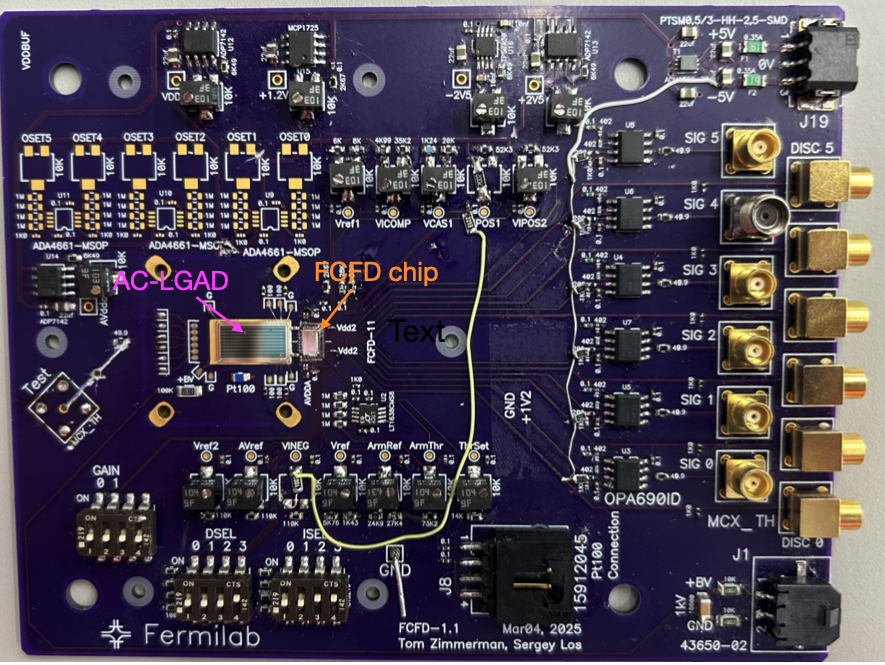}\\
  \caption{A photograph of the readout board hosting the Hamamatsu 1~cm long AC-LGAD strip sensors and the FCFD1.1 chip.}
  \label{fig:ReadoutBoard}
  \end{center}
\end{figure}

The specialized readout board with dimensions $128\times 100$~mm$^2$ shown in Figure~\ref{fig:ReadoutBoard} was designed to help with the evaluation of the FCFD1.1 performance using LGAD sensors of various types.
The mounting pad was designed to hold LGAD sensors of dimensions 12$\times$8~mm$^2$.
The LGAD wire bonding pad structure allows for an equivalent input capacitance connection to the FCFD1.1 channels for debugging purposes. 
In order to minimize wire-bond inductance for the most critical connections, the chip is mounted in a PCB well, flush with the top board surface. 
Such placement shortens the wire bonds to 0.2-0.25~cm for ground, 0.4-0.6~cm for fast signal, and 1.3-1.5~cm for power and current bias connections. 
Microstrip lines are used to route fast signals with an estimated analog bandwidth of about 10~GHz. 
Low voltage power and LGAD bias voltage connections use balun filters to minimize external noise. 
A number of board mounted switches are used to select an amplitude for calibration charge injection and to configure other FCFD1.1 parameters.
A Pt RTD is used to provide monitoring of the board’s temperature.

The readout board mounted with the FCFD1.1 chip and LGAD was installed in a dark environmental chamber for all measurements. 
This chamber is light-tight and allows for full control of the temperature and humidity to ensure reproducible results. 
The output data from the FCFD1.1 chip was recorded using a Lecroy Waverunner 8208HD oscilloscope with an analog bandwidth of 2~GHz and sampling rate of 10~Gsps. 

\subsection{Beam test setup}
The beam test data was collected at the DESY \SI{5}{\GeV} electron beam line and the CERN SPS \SI{120}{\GeV} hadron beam line, using similar LGAD characterization setups described in previous studies~\cite{Apresyan:2020ipp,Heller_2022,Madrid:2022rqw,Dutta:2024ugh}. 
The experimental setup includes a silicon tracking telescope for impact position measurements of each electron/proton, and a Photek 240 MCP-PMT to obtain the reference timestamp for time resolution measurements. The silicon tracking telescope consists of six sensor planes using MIMOSA 26 monolithic active pixel devices with a 18.4~$\mu$m pitch~\cite{Jansen:2016bkd}, all of which were in use during our testbeam campaign.
The MCP-PMT timing response has a resolution of about \SI{10}{ps}~\cite{Ronzhin:2015idh}.
The FCFD1.1 and MCP-PMT waveforms were recorded using the eight channel Lecroy Waverunner 8208HD oscilloscope. 
Since the oscilloscope has only 8 channels, we are able to simultaneously record signals from a maximum of 3 strips: one analog and one discriminator output per strip. 

\begin{figure}[htp]
\centering
\includegraphics[width=0.90\textwidth]{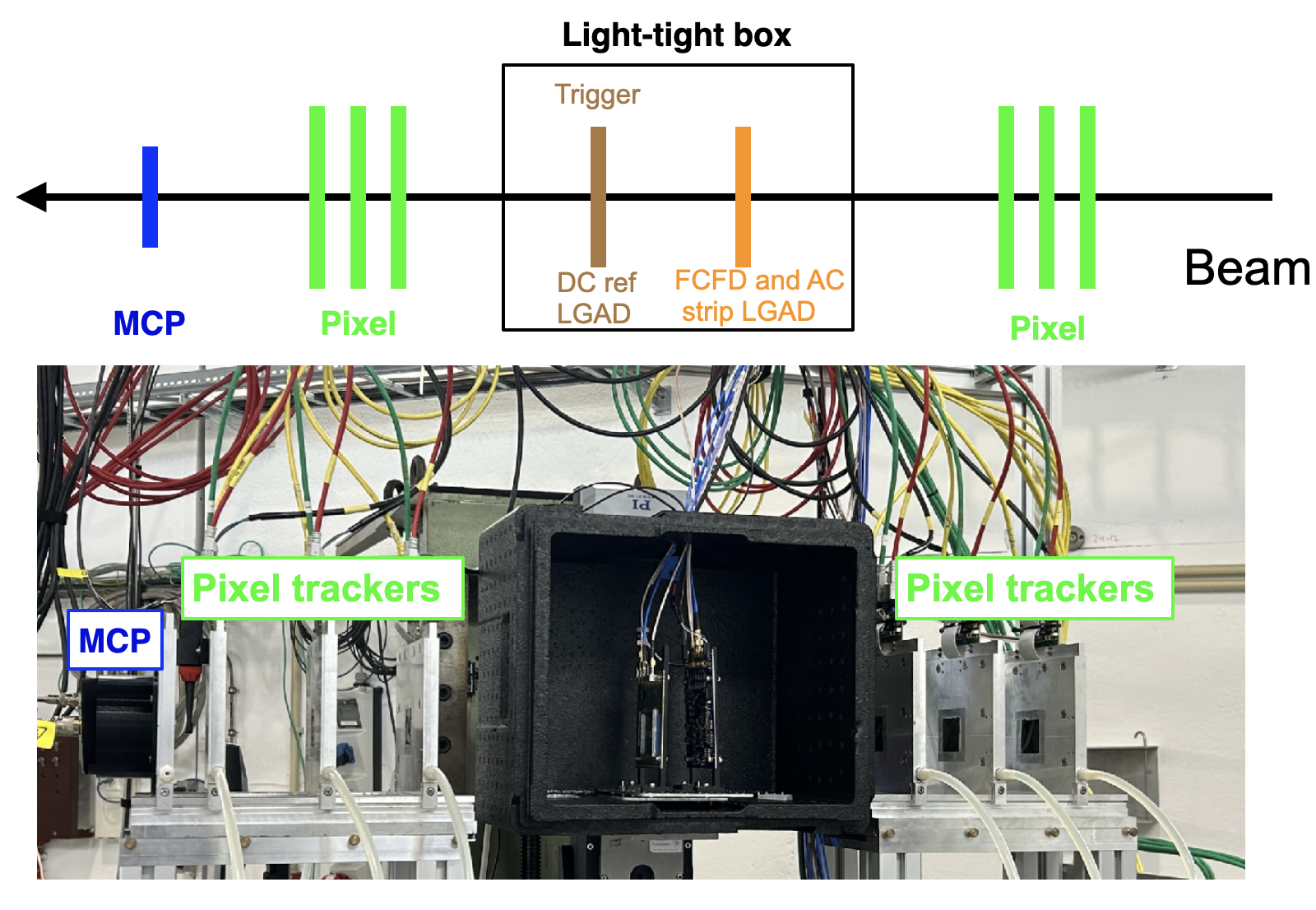}
\caption{Schematic diagram of the DESY setup along the beamline (top), and a photograph (bottom) of the environmental chamber used to hold the FCFD1.1 readout board located in the middle of the tracking telescope.
\label{fig:Testbeam_Box}}
\end{figure}

In Figure~\ref{fig:Testbeam_Box} we show a schematic diagram of the beamline setup and a photograph of the environmental chamber used to contain the FCFD1.1 readout board.
For the CERN test-beam setup, the trigger signal is generated by a scintillator detector located downstream from the environmental chamber and distributed to the tracker telescope and the oscilloscope. 
For the DESY test-beam setup, the trigger signal is generated by the reference DC-LGAD placed inside the same light-tight box. Position resolutions of the impact position achieved with this telescope setup ranges between 5 to 10~$\mu$m depending on the alignment procedure used and the number of tracker planes in use. Events are built by merging the telescope and oscilloscope data offline matching trigger counters from each system. The telescope data are reconstructed with the Corryvreckan software~\cite{Dannheim:2020jlk}.


\section{Results}\label{sec:results}

\begin{figure}[htp]
\centering
\includegraphics[width=0.47\textwidth]{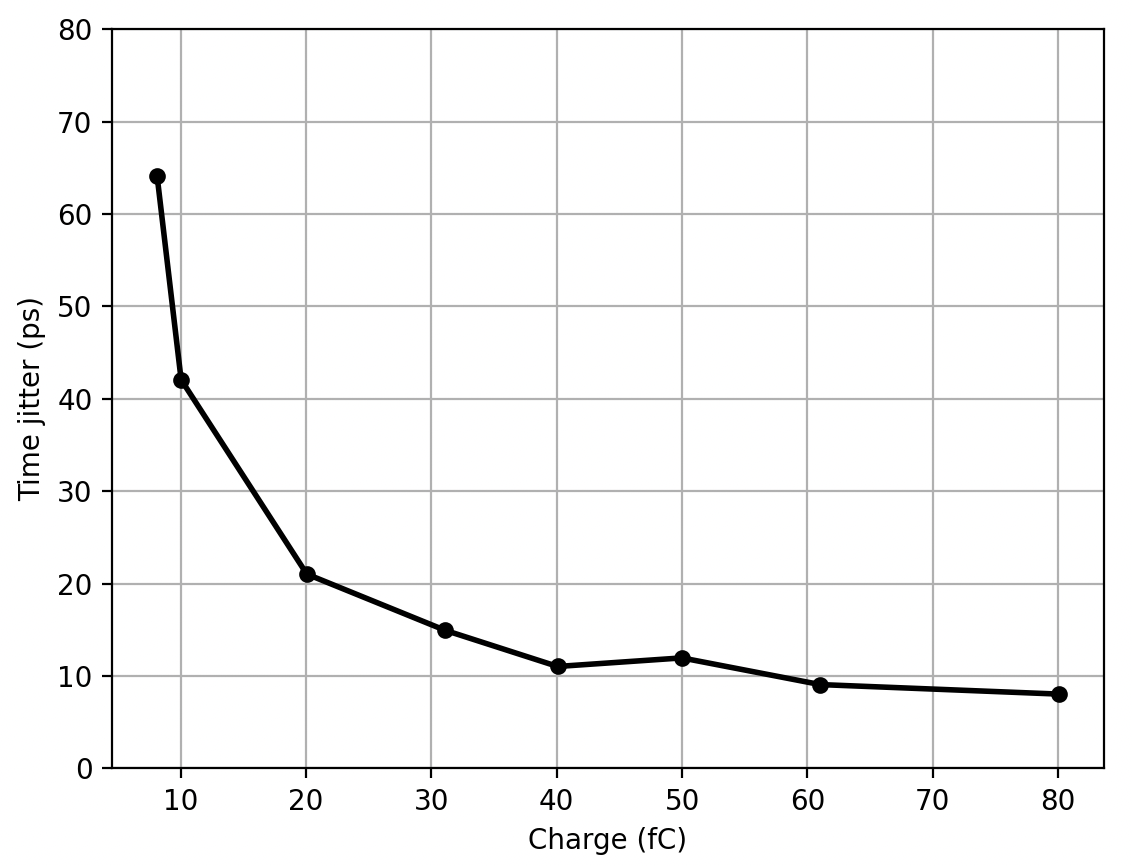}
\includegraphics[width=0.49\textwidth]{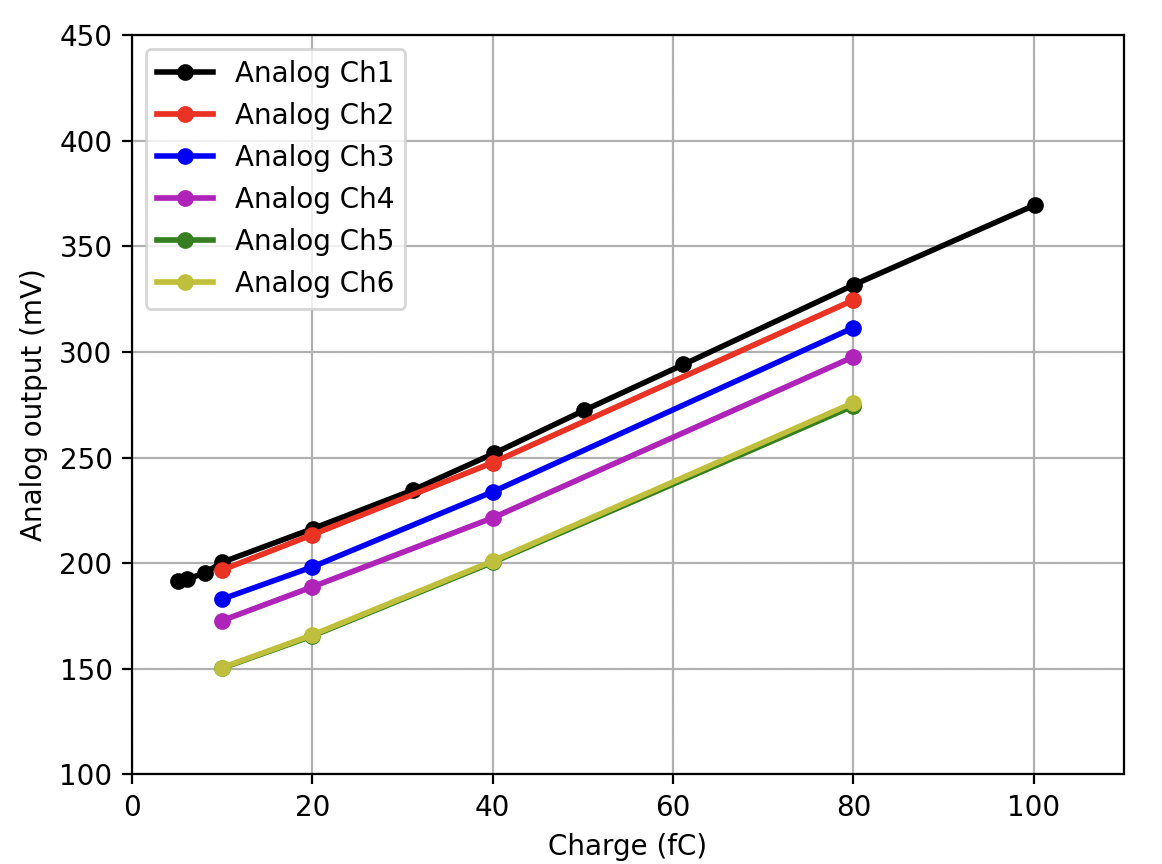}
\caption{Measured time resolution of the FCFD1.1 as a function of the injected charge (left) and the measured amplitude of the analog output channel as a function of the charge injected (right).  
\label{fig:ChargeInjectionResult}}
\end{figure}

We first present characterization results using the internal charge injection mechanism on the FCFD1.1 chip. 
The time resolution (jitter) is measured as a function of the signal size in units of charge, and shown on the left of Figure~\ref{fig:ChargeInjectionResult}. 
We observe results consistent with the measurement from the FCFD0 chip, achieving a resolution better than 20~ps for signals with charge above 20~fC and reaching 10~ps with charge above 40~fC injected. 
The right of Figure~\ref{fig:ChargeInjectionResult} shows the analog output amplitude as a function of the injected charge for all six channels and linear amplitude response is observed for all six channels. 

\subsection{AC-LGAD}

\begin{figure}[htp!]
\centering
\includegraphics[width=0.45\textwidth]
{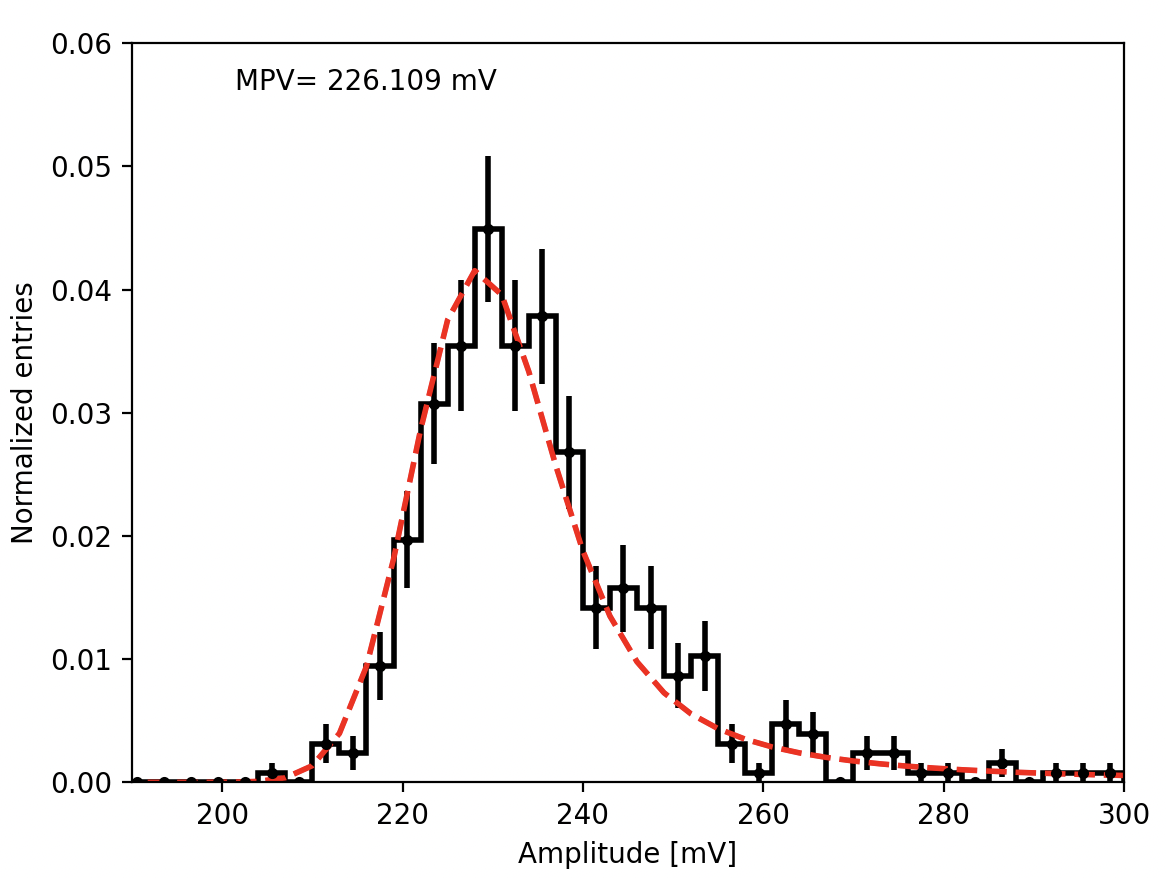} 
\includegraphics[width=0.5\textwidth]{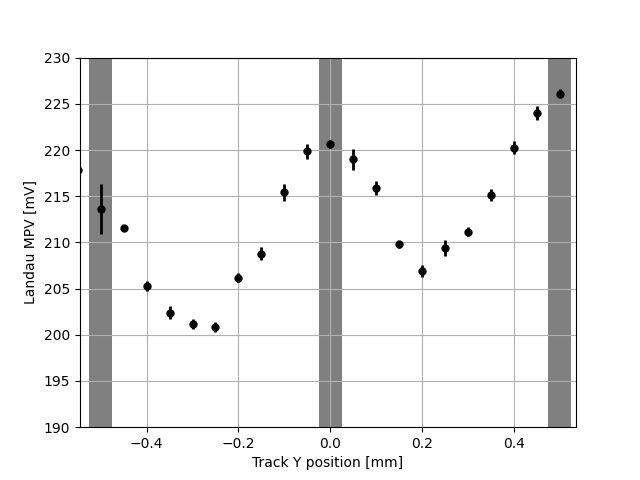} 
\caption{The distribution of the signal amplitude for particles impinging upon the central metal strip (left), and a graph of the fitted Landau MPV parameter as a function of the track position (right). The location of the metal electrodes of the three strips are indicated by the gray bands.
\label{fig:StripAmplitude}}
\end{figure}

Next, we measure the FCFD1.1 chip performance in particle beams with signals generated in the connected AC-LGAD sensors.
We verify the signal amplitude readout functionality of the chip for various channels. 
A histogram of the signal amplitude for one of the AC-LGAD strip channels when the particle directly impacts the metal electrode is shown on the left of Figure~\ref{fig:StripAmplitude} and is fitted to a Landau distribution convoluted with a Gaussian function. 
On the right of Figure~\ref{fig:StripAmplitude} we show the MPV parameter of the Landau distribution obtained from the fits as a function of the track Y position. 
We observe that the signal amplitude is largest when particles hit the metal electrode directly, and gradually decreases as the hit position moves towards the region in between two adjacent strips.

\subsubsection{Analog signal amplitude}
\begin{figure}[htp!]
\centering
\includegraphics[width=0.6\textwidth]{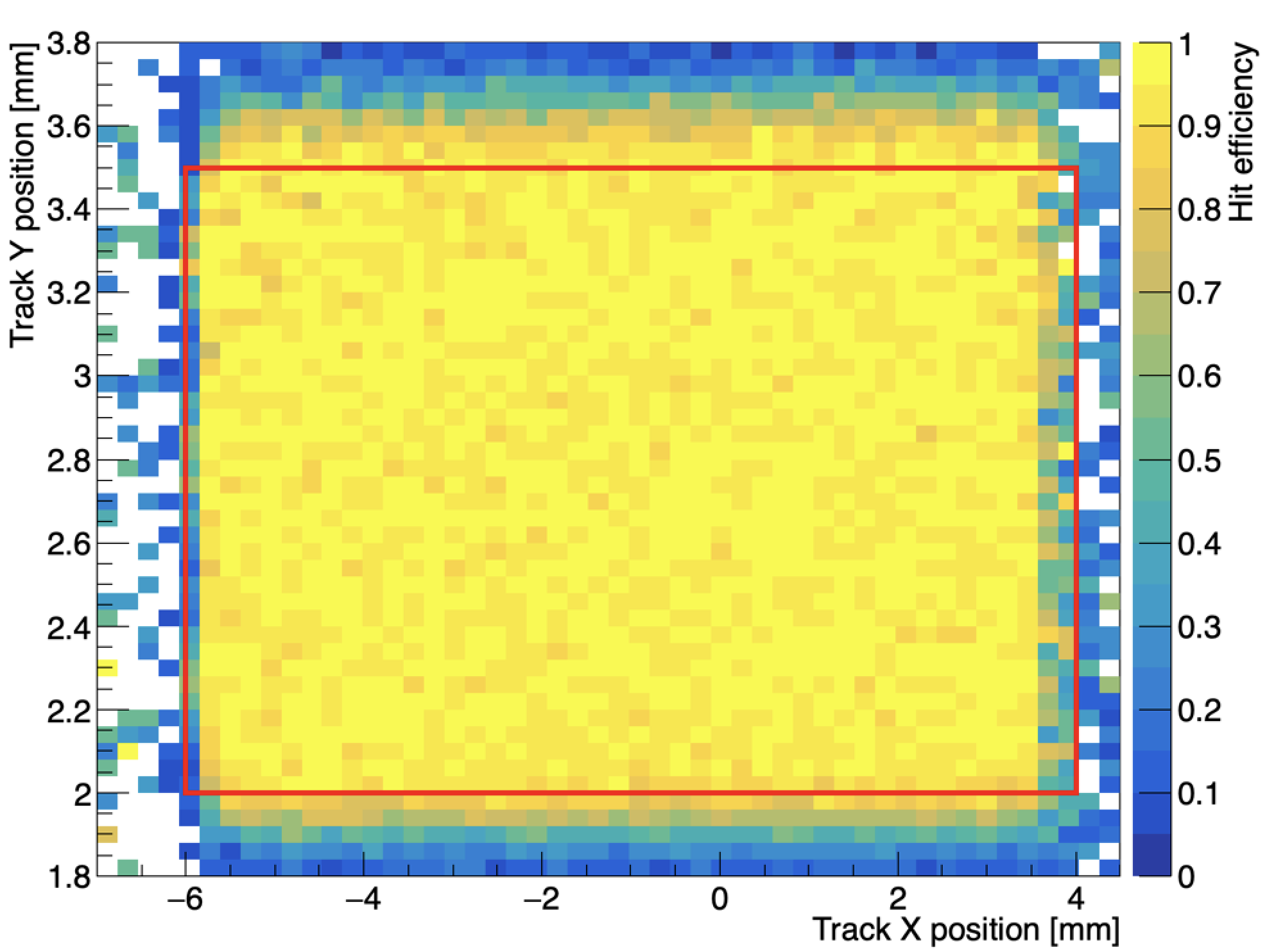}
\caption{The hit detection efficiency of the FCFD1.1 chip as a function of the reconstructed track position. The red rectangle denotes the active area of the strip sensor.  
\label{fig:Efficiency}}
\end{figure}

Using the pixel tracking telescope, we are able to measure the efficiency of the AC-LGAD detection read out by the FCFD1.1 amplitude channel as a function of the X and Y position of the particle impact on the AC-LGAD sensor.
In Figure~\ref{fig:Efficiency}, we present the efficiency of the proton detection in either of three neighboring AC-LGAD strips with a minimum signal amplitude requirement of 190~mV.
We observe that the sensor and FCFD1.1 readout is fully efficient within the 1.5~mm$\times$10~mm active area covered by the three AC-LGAD strips being read out.

\begin{figure}[htp!]
\centering
\includegraphics[width=0.49\textwidth]{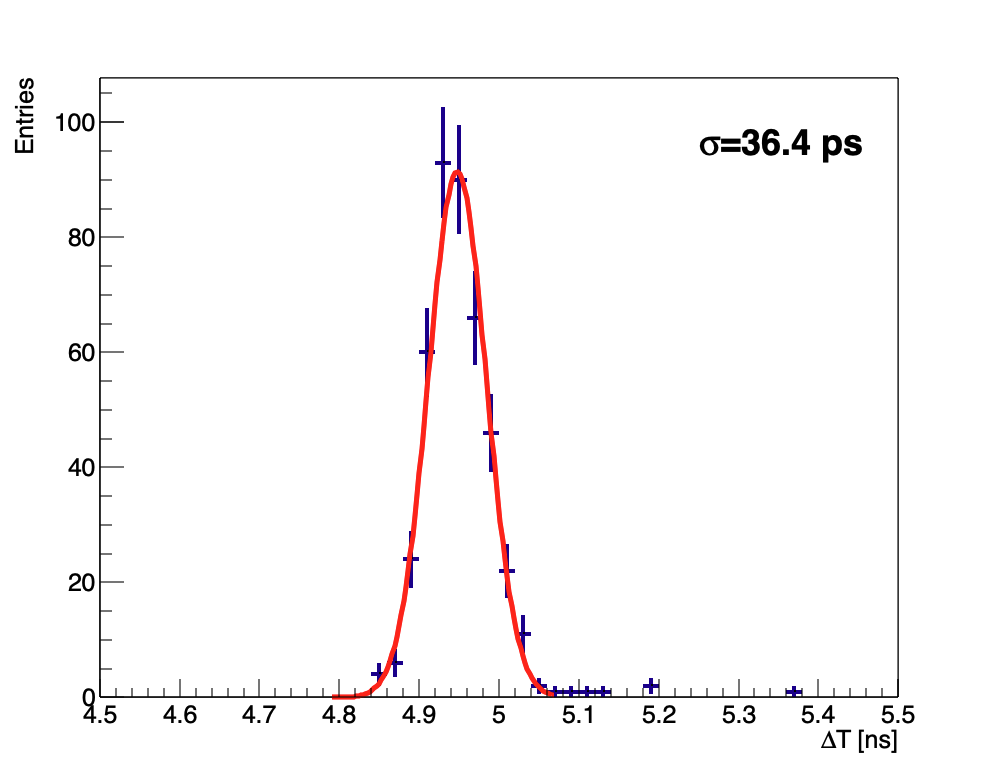}
\includegraphics[width=0.49\textwidth]{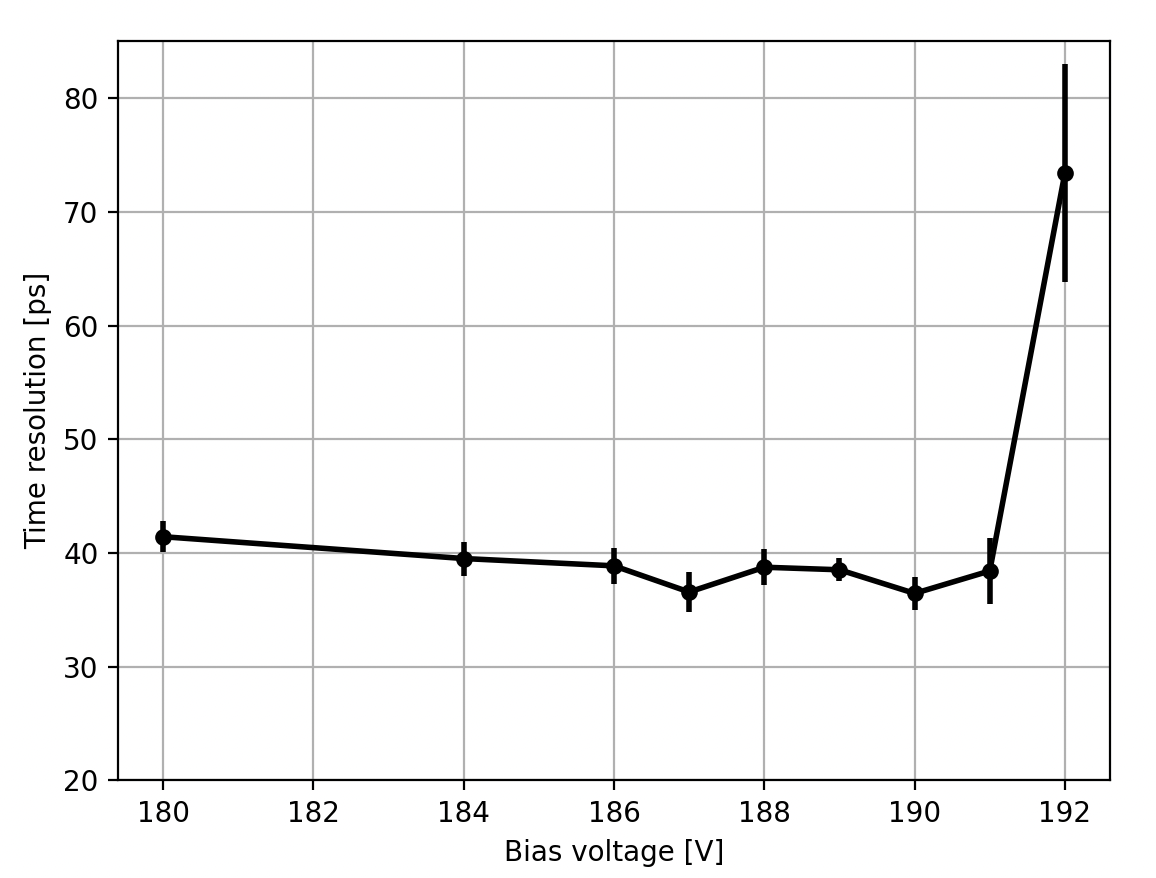}
\caption{Histogram of the $\Delta$t distribution (left) fitted to a Gaussian function used to extract the time resolution parameter ($\sigma$). The time resolution is shown as a function of the bias voltage applied to the AC-LGAD sensors (right). 
\label{fig:ACLGAD_timing_results}}
\end{figure}

\subsubsection{Time resolution}
Next, we characterize the timing performance of the FCFD1.1 chip on the AC-LGAD sensor signal. 
We select events with the track hit position directly on the strip metal electrode of width 50~$\mu$m and within a region of length 1~mm along the strip direction.
The analog channel amplitude of the AC-LGAD is required to be larger than 190~mV and the MCP signal amplitude is required to be above 100~mV. 
In Figure~\ref{fig:ACLGAD_timing_results}, we show the histogram of the time of arrival on the left and the time resolution measurement as a function of the applied bias voltage on the right. 
We achieve a time resolution of 36~ps at a bias voltage setting of 190~V. 
We observe that the time resolution improves with the bias voltage until it reaches above 191~V where it quickly degrades as it approaches breakdown. 
From these measurements we confirm that the achieved time resolution fully meets the design specification.

\begin{figure}[htp!]
\centering
\includegraphics[width=0.8\textwidth]{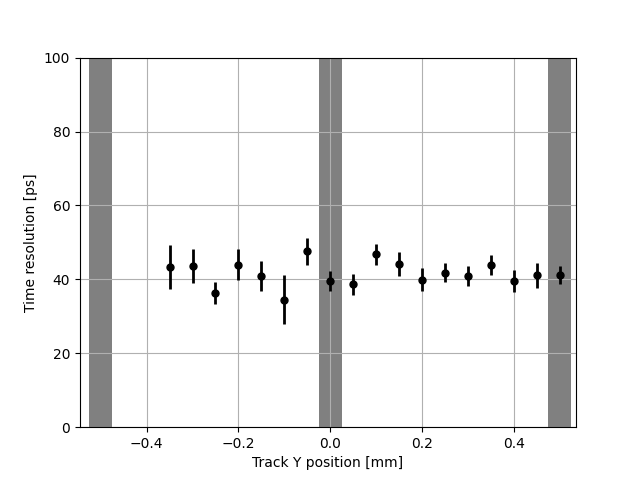}
\caption{AC-LGAD time resolution as a function of track position. Grey areas denote the metalized strip electrodes. The missing data points near the first electrode is due to lack of recorded events with particles impinging in that region.
\label{fig:ACLGAD_timing_vs_Y}}
\end{figure}

Given the large length of these sensors, a position-dependent delay results due to geometric effects.
To correct for this effect, one could use the external tracker to determine the proton hit position as a function of X and Y, and creates a reference map of correction values depending on the location of the hit. 
We employ an alternative correction strategy which utilizes the amplitude of the analog output from the leading and sub-leading channels on the sensor to correct for the delay using the formula~\cite{Dutta:2024ugh}:
\begin{equation}
  t=\frac{a{_1}^2t_{1}+a{_2}^2t_{2}}{a{_1}^2+a{_2}^2}
\end{equation}
where subscripts 1 (2) refers to the leading (subleading) channel, $t_{i}$ is the tracker-corrected time of arrival and $a_{i}$ is the signal amplitude. 
The time resolution measured with the position corrected timestamp as a function of the track Y position is shown in Figure~\ref{fig:ACLGAD_timing_vs_Y}. 
We obtain a fairly uniform resolution of about 40~ps across all three strips. 
Due to the limited event sample size, the statistical uncertainties are larger for the two left-most strips on the plot.

\subsubsection{Spatial resolution}
Spatial reconstruction of the particle hit position benefits from signal sharing between adjacent channels in the AC-LGAD sensors. 
Our position reconstruction strategy employs the selection requirements described above for the time reconstruction, and in addition to selections that determine the feasibility of an interpolated reconstruction using multiple neighboring channels. 
These interpolation techniques were previously introduced as one- and two-strip reconstruction in ~\cite{Dutta:2024ugh}. 
The driving factor for the performance of the spatial resolution of these sensors lies in the efficiency of their two-channel reconstruction for signal proton hits. 
For two-channel position determination, it is derived similarly as the {\it multi-channel timestamp}:
\begin{equation}
  s=\frac{a{_1}^2s_{1}+a{_2}^2s_{2}}{a{_1}^2+a{_2}^2}
\end{equation}

where subscript $1$ ($2$) refers to the leading (subleading) channel: {\it $s_1$($s_2$)} is the actual center position of the leading (subleading) channels and $a_1$($a_2$) is the amplitude of the leading (subleading) channel respectively. 
Near the center of the gap region, the majority of events fall in the two-channel category. 
However, signals from protons that strike very close to the metalized strip fall mostly in the one-channel category, resulting in a larger uncertainty in the measured resolution using the two-strip combination method.

\begin{figure}[htp!]
\centering
\includegraphics[width=0.48\textwidth]{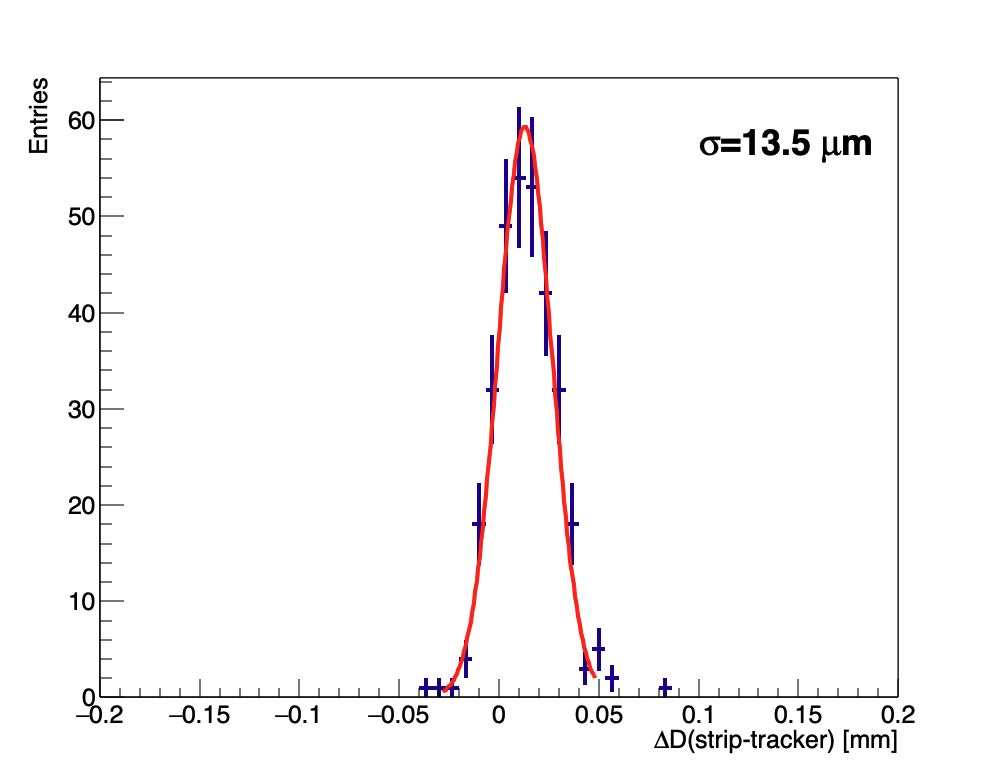}
\includegraphics[width=0.5\textwidth]{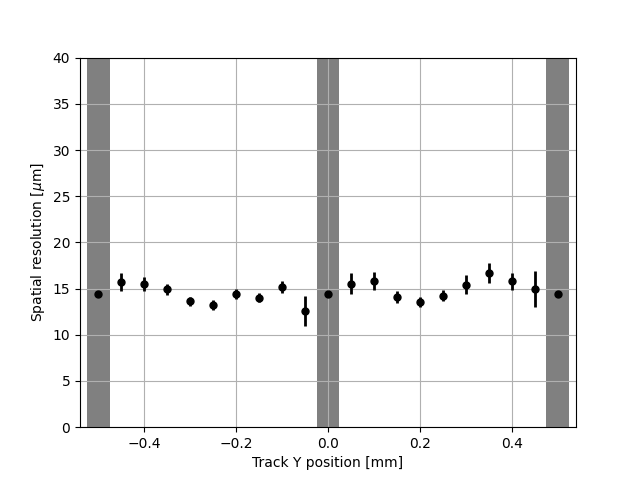}
\caption{Distribution of the residual for the position measurement between the strip sensor under test and the tracking telescope (left). The spatial resolution is shown as a function of the track position (right).
\label{fig:ACLGAD_spatial_vs_Y}}
\end{figure}

On the left of Figure~\ref{fig:ACLGAD_spatial_vs_Y} we show the distribution of the residual for the position measurement between the strip sensor under test and the tracking telescope. 
On the right of Figure~\ref{fig:ACLGAD_spatial_vs_Y} we show the spatial resolution obtained from fitting the distance difference distribution as a function of the tracker $Y$ position. 
When the particles land in the metalized strips, a fixed resolution of 50$\mu$m/$\sqrt{12}$=14.4~$\mu$m is assigned. 
We obtain a fairly uniform spatial resolution around 15~$\mu$m along the channel direction (y). 
Thanks to the analog output, the strip amplitude weighted measurement gives an order of magnitude improvement over the strip only (500~$\mu$m/$\sqrt{12}$=144~$\mu$m).

\section{Conclusions and Outlook}\label{sec:discussion}

We have presented the optimization strategy for the Fermilab Constant Fraction Discriminator (FCFD) readout ASIC and the performance characterization of the new FCFD1.1 prototype. 
A comprehensive evaluation was performed using both injected charge signals and minimum-ionizing particles in test-beam conditions. 
The results demonstrate that FCFD1.1 meets its design specifications, achieving a timing jitter of approximately \SI{10}{ps} for injected charges above 40 fC, a time resolution of \SI{40}{ps} in particle beams, and a position resolution of $15~\mu$m across the 1-cm AC-LGAD strip sensors with a 500~$\mu$m strip pitch. 
Relative to the previous version of the chip~\cite{Xie:2023flv}, FCFD1.1 features a fully redesigned front end, an expanded dynamic range, and a new analog-amplitude readout required for precise position reconstruction. 
We developed and demonstrated a dedicated procedure for extracting the RC-parameters of AC-LGAD strip sensors that are essential to the readout-chip design.

Development of the next FCFD version is currently underway. 
The forthcoming version will serve as the first mixed-sensor prototype for the barrel time-of-flight detector of the ePIC experiment at the EIC. 
It will integrate all components needed for system-level testing and module assembly, including on-chip communication and configuration blocks, ADC and TDC blocks, and an I\textsuperscript{2}C interface. 
These elements, together with the floorplan and power-distribution scheme, will be implemented in a six-channel prototype as a precursor to the full-scale readout ASIC.

\section*{Acknowledgements}

We thank DESY and CERN for providing the test beam for the excellent performance of the accelerator and support of the test beam facility, in particular A.~Herkert, M.~Stanitzki, S.~Ackermann, M.~Doser, M.~Jaekel, L.~Zwalinski and M. Van Dijk. We also thank A.~Rummler for help with the telescope tracker system and providing late night supports.

We also thank the SiDet department, in particular M.~McDonough and H.~Gonzalez, for preparing the readout board by mounting and wirebonding the LGAD sensor along with the FCFD1.1 ASIC. 

This document was prepared using the resources of the Fermi National Accelerator Laboratory (Fermilab), a U.S. Department of Energy, Office of Science, HEP User Facility. 
Fermilab is managed by Fermi Research Alliance, LLC (FRA), acting under Contract No. DE-AC02-07CH11359.
This work has also been supported by funding from the California Institute of Technology High Energy Physics under Contract DE-SC0011925 with the U.S. Department of Energy. 
This manuscript has been authored by an author at Lawrence Berkeley National Laboratory under Contract No. DE-AC02-05CH11231 with the U.S. Department of Energy.

\bibliographystyle{elsarticle-num} 
\bibliography{main}{}

\end{document}